# Procedure to produce discretized path composed by semi-circle and straigth sub-paths

Sparisoma Viridi

Nuclear Physics and Biophysics Research Division, Institut Teknologi Bandung, Bandung 40132, Indonesia
dudung@fi.itb.ac.id

Some problems founds in teaching physics related to curved paths that are unfortunately only described as illustration. A simple way to introduce the path is presented, which can help students to test their concept numerically. The procedure is limited into semi-circle and straight sub-paths. Smaller discretizing width $\Delta s$ gives better form of the produced path.

In some problems such as mechanics [1], thermodynamics [2], fluids [3], and electrostatics [4], information about path is required, but it is sometimes avoided by introducing other higher level and more sophysticated concepts. The reason is simply to avoid mathematical difficulties. Illustration of a path is given in Figure 1.

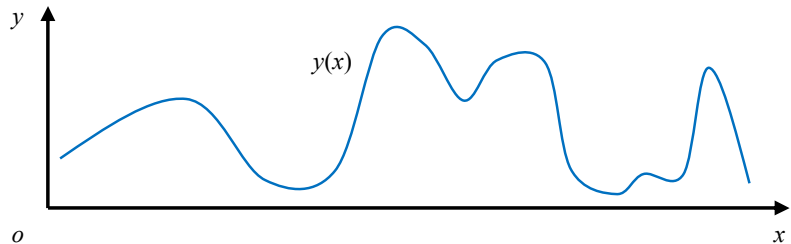

Figure 1. Illustration of path of $y(x)$.

Suppose that an object moves along this path which has friction and the total work done by the friction is to be found. Without information about the path and how the object orientation in every point, the work can not be calculated. It is only an illustration the need to have information about path.

A path $s$ can be discretized into $ds$, which consists of $dx$ and $dy$ as in the formula

$$(ds)^2 = (dx)^2 + (dy)^2, \qquad (1)$$

which can be expressed also in the form of

$$dx = ds \cos\theta, \qquad (2)$$

$$dy = ds \sin\theta, \qquad (3)$$

with $\theta$ is the incline angle of $ds$ measured from $x$ axis as illustrated in Figure 2.

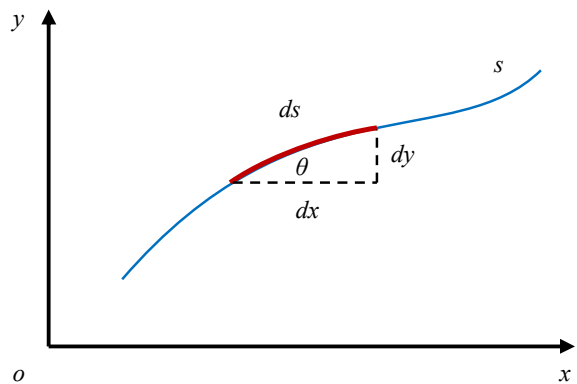

Figure 2. Discretized path $ds$ and its relation with $dx$ and $dy$.

Next step is how to discretize $s$ into $N$ segments with width $ds$. Other example to discretize $s$ is using parametric equation, but it does not give equal $ds$ as reported in disretizing self-siphon path [4]. A procedure to produce equal $ds$ is described in this article.

A simple way to have $ds$ is through

$$ds \approx \Delta s = \frac{s}{N}, \qquad (4)$$

or it can be found that

$$s = \int ds \approx \sum_{i=1}^{N} \Delta s, \qquad (5)$$

for large number of $N$. Equation (1)-(3) can be expressed in vector form which is

$$\vec{ds} = \hat{e}_x dx + \hat{e}_y dy = (\hat{e}_x \cos\theta + \hat{e}_y \sin\theta) ds. \qquad (6)$$

The incline angle $\theta$ can be used as parameter to generate $s$ (explicitely, $x$ and $y$) based on value of $\Delta s$ and iteration index $i$. It can be written that

$$x_{i+1} = x_i + \Delta x_i = x_i + \Delta s \cos\theta_i, \qquad (7)$$

$$y_{i+1} = y_i + \Delta y_i = y_i + \Delta s \sin\theta_i. \qquad (8)$$

Then, discretization of incline angle $\theta$ into $d\theta$ is needed. It can be obtained through

$$d\theta \approx \Delta\theta = \frac{\theta_b - \theta_a}{M}, \qquad (9)$$

with initial angle $\theta_a$ and final angle $\theta_b$. For a straight line it will be found that $\Delta\theta = 0$. In Equation (9) number of segment $M$ is used instead of $N$ as in Equation (4). In Equation (4) value of $N$ is set and $\Delta s$ is obtained. But in Equation (9) value of $M$ is determined throung chosen value of $\Delta s$. The later scheme is preferred in the implementation of discritizing the path $s$. For a semi-circle path with radius $R$ it can be found that

$$\Delta\theta = \frac{\theta_b - \theta_a}{2(\theta_b - \theta_a)R/\Delta s} = \frac{\Delta s}{2}. \qquad (10)$$

Using $\Delta\theta$ it can written that

$$\theta_{i+1} = \theta_i + \Delta\theta, \qquad (11)$$

which holds for several types of sub-path, such as semi-circle and straight path

As an example a path consists of three types of sub-path is presented. Suppose there is a path which consists of a horizontal path with length $L_1$, a semi-circle with radius $R_2$ from $\theta_{2a} = 0$ to $\theta_{2b} = \frac{\pi}{4}$, and incline line with length $L_3$ and angle $\theta_3 = \frac{\pi}{4}$. Then, length of the second sub-path is $L_2 = |\theta_{2b} - \theta_{2a}|R_2 = \frac{\pi}{4}R_2$.

Other way than Equation (4) and (10), is first by defining $\Delta s$ to find $N$ or $M$. This approach will be used instead of the first one. Then, the procedure can be described as follow.

**Step 01**. Start.

**Step 02**. Define value of $\Delta s$.

**Step 03**. Define number of sub-path $M = 3$.

**Step 04**. Start with $j = 1$.

**Step 05**. Define initial angle $\theta_{ja}$, final angle $\theta_{jb}$, and radius $R_j$ of path $j$. (*)

**Step 06**. Define length $L_j$ of path $j$.

**Step 07**. Calculate number of segment $N_j = \dfrac{L_j}{\Delta z}$ of path $j$.

**Step 08**. Calculate discretized incline angle $\Delta \theta_j = \dfrac{\Delta z}{R}$ of path $j$. (**)

**Step 09**. Increment $j$ value with 1.

**Step 10**. If $j < J$ then goto Step 05.

**Step 11**. Define initial position $x_0$ and $y_0$, also initial angle $\theta_0$.

**Step 12**. Calculate number of segment of all sub-paths $N = \sum_{j=1}^{J} N_j$.

**Step 13**. Start with $i = 0$.

**Step 14**. Calculate $x_{i+1} = x_i + \Delta s \cos \theta_i$.

**Step 15**. Calculate $y_{i+1} = y_i + \Delta s \sin \theta_i$.

**Step 16**. Determine incline angle group $j = 1 + \sum_{k=1}^{3} i \setminus N_k$. (***)

**Step 17**. Assign discretized incline angle $\Delta \theta_i = \Delta \theta_j$ for sub-path $j$.

**Step 18**. Calculate $\theta_{i+1} = \theta_i + \Delta \theta_i$.

**Step 19**. Increment $i$ value with 1.

**Step 20**. If $i < N$ then goto Step 14.

**Step 21**. Finish.

(*) straight line has $R = \infty$, (**) straight line has $\Delta \theta = 0$, (***) $\setminus$ is integer division.

Result of the procedure for given problem is as illustrated in Figure 3. It can be seen from the figure that smaller value of $\Delta s$ gives better curve of path $s$.

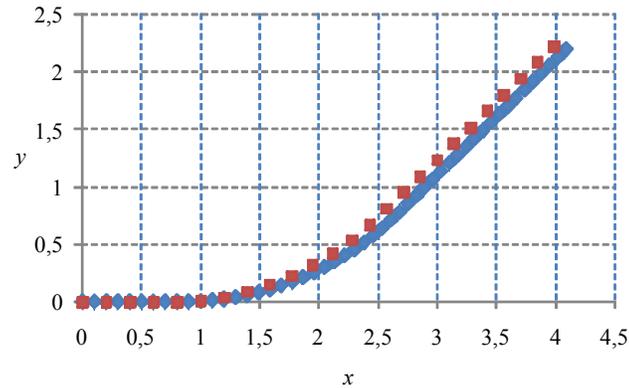

Figure 3. Result of discretized path for the mentioned example for different $\Delta s$ : 0.2 (■) and 0.1 (◆).

For the given example value of $\Delta s = 0.1$ already gives a nearly good results. The incline angle $\theta_{3a} = \theta_{3b} = \dfrac{\pi}{4}$ can be verified from the grid in the Figure 3 that the third sub-path nearly paralel to the diagonal line in each grid. This incline angle is also observed for $\Delta s = 0.2$, but the last sub-path is not so coincided with the last sub-path for $\Delta s = 0.1$. The difference is due to the chosen value of $\Delta s$. Further modification of given procedure can be use to interpolate every curved path.

Back to paragraphs in the beginning of this article, since the path is already discritized then the work done by friction can be calculated numerically. In a segment $i$ with width $\Delta s$ with incline angle $\theta_i$ the friction force would be

$$f_i = \mu_k mg \cos\theta_i, \tag{12}$$

then the work should be

$$W_{f,i} = f_i \Delta z = \mu_k mg \cos\theta_i \Delta z. \tag{13}$$

By summing work done by friction for all segments

$$W_f = \sum_{i=1}^{N} W_{f,i} \tag{14}$$

total work done by friction foce can be calculated. But if the centripetal force is also considered than a correction must be introduced into Equation (12)

$$f_i = \mu_k m \left[ g\cos\theta_i + \frac{1}{R_i}\left(\frac{\Delta z}{\Delta t}\right)^2 \right], \tag{15}$$

with $\Delta t$ is time needed to elapse the segment $i$ with width $\Delta s$. For a straigth line then the second term inside the rectangular bracket in the right side of Equation (15) would be zero and return to Equation (12). Further detail of implementation of Equation (15) is out of the scope of this article.

Procedure to produce dicritized path composed by straight and semi-circle sub-paths has been presented and the result has been shown. Smaller value of $\Delta s$ gives better result to form of the produced path.